# The Integral Field Spectrograph of SPHERE: the Planet Finder for VLT.

R.U. Claudi[1], M. Turatto[1], J. Antichi[1], R. Gratton[1], S. Scuderi[2], E. Cascone[3], D. Mesa[1], S. Desidera[1], A. Baruffolo[1], A. Berton[1,4], P. Bagnara[1], E. Giro[1], P. Bruno[2], D. Fantinel[1], J.-L. Beuzit[5] , P. Puget[6], K. Dohlen[7].

[1] INAF, Astronomical Observatory of Padova, vicolo Osservatorio, 5 35122 Padova Italy;
[2] INAF, Astronomical Observatory of Catania, via S. Sofia 78, 95123, Catania Italy;
[3] INAF, Astronomical Observatory of Napoli,Salita Moiariello, 16, 80131, Napoli, Italy;
[4] Max Planck Institute for Astronomy, Königstuhl 17, D - 69117 Heidelberg, Germany;
[5] Laboratoire d'Astrophysique Observatoire de Grenoble BP 53X, 414 rue de la Piscine F-38041 Grenoble cedex 9 France
[6] Observatoire de Paris, 5, Place Jules Janssen 92195 Meudon Cedex (France)
[7] LAM - Laboratoire d'Astrophysique de Marseille, 2 place Le Verrier13248 Marseille cedex 4.

## ABSTRACT

SPHERE is an instrument designed and built by a consortium of French, German, Italian, Swiss and Dutch institutes in collaboration with ESO. The project is currently in its Phase B. The main goal of SPHERE is to gain at least one order of magnitude with respect to the present VLT AO facility (NACO) in the direct detection of faint objects very close to a bright star, especially giant extrasolar planets. Apart from a high Strehl ratio, the instrument will be designed to reduce the scattered light of the central bright star and subtract the residual speckle halo. Sophisticated post-AO capabilities are needed to provide maximum detectivity and possibly physical data on the putative planets.
The Integral Field Spectrograph (IFS), one of the three scientific channels foreseen in the SPHERE design, is a very low resolution spectrograph (R~20) which works in the near IR (0.95-1.35 μm), an ideal wavelength range for the ground based detection of planetary features. Its goal is to suppress speckle to a contrast of $10^7$, with a goal of $10^8$, and at the same time provide spectral information in a field of view of about 1.5 × 1.5 arcsecs$^2$ in proximity of the target star.
In this paper we describe the overall IFS design concept.

## 1. INTRODUCTION

SPHERE (Spectro-Polarimetric High-contrast Exoplanets REsearch:) is a second generation instrument for VLT optimized for high contrast imaging at the Nasmyth focus of one of the VLT units. The design of SPHERE is divided into four subsystems: the Common Path Optics and three science channels, a differential imaging camera (IRDIS), an Integral Field Spectrograph (IFS), and a visible imaging polarimeter (ZIMPOL). The Common Path includes pupil stabilizing fore optics (tip-tilt and derotator) where insertable polarimetric half-wave plates are also provided, the SAXO extreme adaptive optics system with a visible wavefront sensor, and NIR coronagraphic devices in order to feed IRDIS and IFS with a highly stable coronagraphic image in the NIR. ZIMPOL shares the visible channel with the wavefront sensor and includes its own coronagraphic system. While the ZIMPOL arm compares observations on two orthogonal polarization in order to detect the polarized light reflected by the planet on the stellar background, the former two channels (IRDIS and IFS) explore the stellar neighbourhood in order to find planetary spectral features. For the IFS, this quest is conducted searching for strong $CH_4$ absorption bands in both the stellar light reflected by gaseous Jupiter-like planets and in thermal emission from young-warm planets. Additional science topics which will be addressed by SPHERE include the study of brown dwarfs, circumstellar disks and related phenomena such as mass loss mechanisms, stellar winds, planetary nebulae and perhaps even supernova explosions.

## 2. INSTRUMENT SPECIFICATION AND CONCEPTUAL DESIGN

The limiting factor for planet detection with adaptive optics is the speckle noise. This can be greatly reduced by differential techniques such as methane band differential imaging. This approach takes advantage of the large variation of the planet brightness across methane bands (or other molecular features). The subtraction of an image at the wavelength of an absorption band from an image at a continuum wavelength eliminates the speckles to first order, but leaves the planet signal intact. The wavelength dependence of the speckles set the limits on the noise reduction achievable in this way, but it has been shown that double difference schemes employing images at three wavelengths can be very effective[1]. These methods can be implemented via software if a data cube is available. It is, in fact, possible to refine this approach by searching for planets in a data cube with a number of different schemes that optimize the detectability for a set of planet template spectra[2].

The basic concept of the SPHERE Integral Field Spectrograph is based on TIGER[3] and SAURON[4] designs. The scientific case imposes the following basic requirements for the IFS: a minimum spectral coverage between 0.95 and 1.35μm, with a very low resolution (R=20@1.15 μm; the disperser used to achieve this kind of resolution is discussed elsewhere[5]); a minimum FOV of 1.5 arcsec × 1.5 arcsec (2.12 arcsec diagonal) and a high throughput in order to achieve high contrast requirements. The camera mounts an IR 2k × 2k detector.

In order to determine the spatial sampling we assume that the adaptive optics module located before the Integral Field Units (IFU) concentrates a large fraction of light from the planet within the first diffraction peak of the telescope, the FWHM of which, at the middle of the spectral range, is 29.6 mas. Currently we favor a spatial sampling of 12.3 mas per microlens as a good trade off between high angular resolution and signal to noise ratio of the spectra.

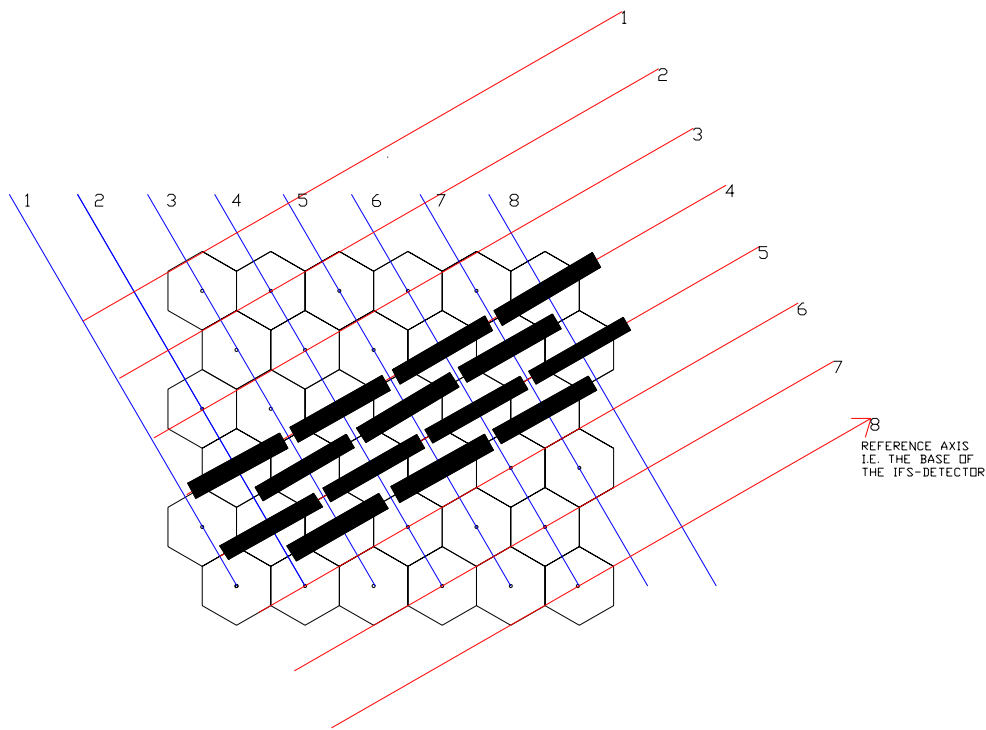

**Figure 1: Thin hexagons represent the microlens array (IFU). Solid rectangles represent the arrangement of the spectra on the IFS focal plane. Oblique rows and columns represent the orientation of the IFS Detector**

Since we aim to detect light emitted or reflected directly from extrasolar planets, we have to cover as much as possible the field where these objects are likely to be detectable. The FoV as defined by scientific arguments is at least 1.5 × 1.5 arcsec$^2$, and this value, coupled with the spatial sampling determined above, constrains the array size to 122×122 lenses. The lens size is a function of the spectral parameters (length and separation) and of the geometrical arrangement of the lenses-array. On one hand we need a configuration that guarantees that the distance between neighbour pixels is adequate to avoid cross talk contamination; on the other hand each single spectrum has to include the whole spectral range without overlapping in the direction of the dispersion.

Fulfillment of these two conditions allows the best exploitation of the detector area, without denying the requirements imposed by SPHERE science goals.

The presence of adjacent spectra could be neglected in data reduction[6] for a crosstalk level of $10^{-4}$. However due to the presence of the wings of the diffraction figure in the spatial direction, (orthogonal to the dispersion direction of each spectra), a realistic crosstalk level is about $10^{-3}$ for the whole spectral range. In the configuration we present in Sect. 3, this value may be obtained adopting a separation between columns of adjacent spectra of 4.17 pixels. In this case the crosstalk between adjacent spectra cannot be neglected in the spectrum extraction procedure; the crosstalk should then be measured by appropriate calibrations, and should be taken into account in the data reduction by a suitable procedure.

The spectral length is a function of the spectral range and of the resolution of the disperser. In our case we evaluate a spectral length $l_S$ =21.4 pixels.

The allowed configuration of the spectra on the IFS detector focal plane depends on the lenses shape. We have found the best solution for SPHERE IFS considering both hexagonal and square shapes, which represent the best choices in terms of filling factor. We have investigated a sub sample of configurations that keep distance ($\delta_S$) constant along the whole length of adjacent spectra in order to find the best ones given by crosstalk and resolution arguments. The best configuration is obtained adopting the hexagonal lenses-array IFU shown in Figure 1.

Considering the geometrical and spectral parameters described above and the preliminary IFS optical design described in the following paragraph, the size of the IFU will be of 26 ×26 mm$^2$, adopting a diameter of 150 μm for the individual lenslet.

## 3. TOWARDS AN OPTICAL DESIGN OPTIMIZED FOR PLANETS DETECTION

This section describes the optical constraints that an IFS based on TIGER and SAURON optical designs should verifies in order to allow that the Spectroscopic Simultaneous Differential Imaging Technique (hereafter SSDI), resumed in the section 2, applies correctly to the science case.

### 3.1. IMAGE QUALITY EFFECTS ON DIFFRACTION LIMITED INTEGRAL FIELD SPECTROSCOPY

In applications of integral field spectroscopy to diffraction limited cases, three peculiar effects due to diffraction are important and should be taken into account for an optical design based on lens-array:

(i) Spectrum-to-spectrum cross talk
(ii) Variation of the lens-array diffraction pattern due to not uniform illumination of the lens-array
(iii) Speckle chromatism

### 3.1.1. Spectrum-to-spectrum cross talk

In general, the monochromatic total signal $S(\lambda)$ from a point in the field of view will be:

$$S(\lambda) = s(\lambda) + p(\lambda) + \int_{\lambda_{min}}^{\lambda_{max}} c(\lambda')S(\lambda')d\lambda'$$

where $s(\lambda)$ is the monochromatic signal of the parent star, $p(\lambda)$ is the monochromatic signal of the planet, $\lambda_{min}$ ($\lambda_{max}$) is the minimum (maximum) wavelength of the spectrum. Finally, $c(\lambda)$ is the monochromatic cross-talk of the nearest neighbour spectra onto the reference one at the position corresponding to the wavelength $\lambda$.

Then, we can consider the total signal $S(\lambda)$ at two wavelengths one with and one without the feature due to the putative planet. The monochromatic signals are:

$S(\lambda_1) = s(\lambda_1) + p(\lambda_1)$
$S(\lambda_2) = s(\lambda_2) + p(\lambda_2)$

Basic assumptions of SSDI are: $p(\lambda_2)=0$ and $s(\lambda_2)= s(\lambda_1)$.

It is possible to estimate the monochromatic star/planet contrast in a fixed point inside the FOV (i.e. local monochromatic contrast) around the star at the wavelength affected by the planet feature, by exploiting the following relation:

$$\frac{s(\lambda_1)}{p(\lambda_1)} = \left(\frac{1-SR(\lambda_1)}{SR(\lambda_1)}\right) \times \frac{I_{star}}{I_{planet}} \times \left(\frac{N_{Im}^P}{N_{Im}^S}\right)^2$$

This one expresses the local monochromatic contrast in terms of the monochromatic Strehl Ratio level (SR) of the total signal, in term of the global star-planet contrast ($I_{star}/I_{planet}$) and in term of the ratio between the number of spatial pixels on which the planet signal ($N^P_{Im}$) and the star signal ($N^S_{Im}$) are focused respectively. Then, for a start-planet contrast upper limit $I_{star}/I_{planet} = 10^8$, the polychromatic cross-talk upper limit will be:

$$C = \int_{\lambda_{min}}^{\lambda_{max}} c(\lambda) d\lambda \leq 10^{-4}$$

### 3.1.2. Variation of the lenses-array diffraction pattern due to not uniform illumination

This is justified by considering that each lens generates a diffraction pattern on the lenses-pupil plane that finally is re-imaged onto the detector as in TIGER and SAURON instruments. This optical behaviour is produced by the Nyquist sampling of the telescope Point Spread Function which in turn implies the sub-sampling of the telescope Optical Transfer Function on the lenses-pupil-plane i.e. the entrance-slits-plane of the IFS. Moreover, a perfect Airy-disk pattern for the pupil generated by the single aperture will result only in the case of uniform illumination of the whole lenses-array. This is not the general case and if the lens pitch is such that the lenses-array samples at the Nyquist limit the telescope Point Spread Function, the typical Speckle pattern residual incoming on the lenses-array will produce a percentage of shape-distorted pupils on the lenses pupil plane.

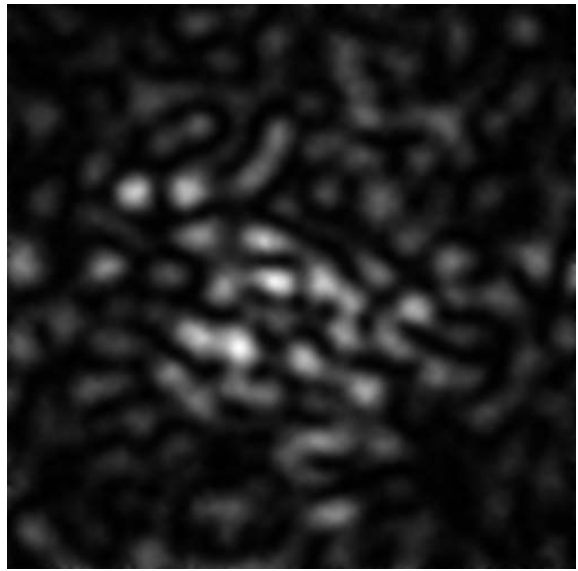

**Figure 2: Example of simulated speckle pattern residual (0.5 msec integration time) after Adaptive Optics correction on a lenses-array with specifications matching the ones described in the text.**

For an optimized IFS design as the one described in section 3.2, the impact of a not uniform illumination on the single pupil is to increase the wings of its diffraction pattern and to decrease the intensity of its central peak: Moreover, the shape variation of the central peak will fall within a single detector pixel, then totally under-sampled.

### 3.1.3. Speckle chromatism

Speckle chromatism is the chromatic dependence of speckles[7] and should be addressed for any kind of IFS working at the diffraction limit condition and exploiting the dispersion of the spectrum information. For such instruments, when a given spectrum is considered on the detector plane, the speckles are seen as oscillations of the spectrum along the dispersion axis.

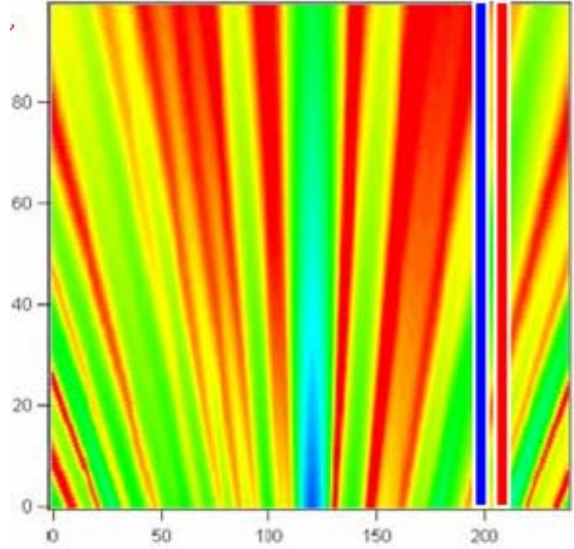

**Figure 3: Experimental slice through a 3D data cube corresponding to a speckle pattern obtained with an ideal Lyot Coronagraph[8]. The y-axis is the one assigned to λ-dispersion while the x-axis corresponds to the radial coordinate assigned on sky.**

As the above Figure 3 shows, the chromatic speckles pattern along the λ-dispersion axis can be represented by the following pseudo-periodic signal:

$$I_{SPECKLE}(\lambda, N) = \cos\{2\pi \cdot [N/\Delta] \cdot \lambda\}$$

where ($\Delta$) is proportional to the wavelength ($\lambda$) and (N) is the radial coordinate ($\rho$) expressed in terms of $\lambda/D$ (D = 8000 mm is the diameter of the VLT entrance pupil at the Nasmyth focus):

$$N = \rho \cdot \frac{\lambda}{D}$$

Is then clear that at a fixed position ($\rho$), the pseudo-period of the spectrum becomes smaller than a fixed spectral resolution ($\Delta\lambda$) for small enough values of the wavelength. This fact produces the chromatic speckle under-sampling (i.e. aliasing effect on the subsequent spectrum re-construction procedure). Moreover the aliasing effect becomes greater when we increase the radial coordinate ($\rho$).

The typical period of spectral speckles ($\Delta/N$) depends on the inverse of the distance from the field centre. Then, there is an inner region where these variations are properly sampled (and interpolation/re-binning can be done safely), and an outer region where this is not anymore possible.

The separation between these two regions of the field is obtained by applying the Nyquist criterion to the spectral speckles. This method allows expressing the radial coordinate threshold ($\rho_{NY}$) in term of the minimum and the central wavelengths $\lambda_{min}$ & $\lambda_c$ of the range covered by the IFS, spectral resolution R and entrance pupil diameter (D):

$$\rho_{NY} = \frac{\lambda_{min}^2 \cdot R}{D \cdot \lambda_c}$$

It is then clear that any IFS working at the diffraction limit condition and exploiting the dispersion of the spectrum information gets a safe region where the spectra are SUPER-sampled, then avoiding aliasing effects before the image acquisition.

Finally, the condition of spectral speckles SUPER-sampling should be coupled with the cross-talk and the not uniform illumination constrains of the lenses-array at the entrance of the IFS, in order to fix the best parameters of a TIGER (or SAURON)-like instrument optimized for planets detection. We define this new IFS optical concept: SUPER-TIGRE.

### 3.2. SUPER-TIGRE OPTICAL DESIGN BASIC PARAMETERS

In order to cover the maximum FOV around the central target star achievable with a 2kx2k-detector (18 μm/pixel) in the spectral range 0.95-1.35 μm, and verifying the following constraint:

- cross-talk upper limit $< 10^{-4}$
- minimum variation of the lenses-array diffraction pattern due to not uniform illumination
- chromatic speckles SUPER-sampling

we obtained the following IFS specifications:
- Maximum FOV= 1.5 x 1.5 arcsec$^2$
- Spectral resolution: R~ 20
- Total number of spectra: 14,884 (122x122 IFU-lenses)
- Pitch of the single IFU-lens: $D_L$= 150 μm
- Focal ratio of the single IFU-lens: $FR_L$ = 16
- Fore-optics focal ratio: $FR_{FORE-OPTICS}$ = 316
- IFS entrance-slits scale: 0.0813 arcsec/mm
- Camera vs collimator magnification: $m_{IFS}$ = 2.16

### 3.3. SUPER-TIGRE OPTICAL DESIGN RAY-TRACING SCHEME

This section describes the optimization scheme we followed to design the IFU+IFS optical system. This latter is a standard on-axis dioptric-train conceived with the following optical scheme:
- IFU-lenses: stops plane
- IFU-pupils i.e. IFS entrance-slits: object plane
- IFS collimator lenses: 1 cemented doublet + 1 cemented triplet
- IFS camera lenses: 2 cemented doublets + 1 singlet + a field lens
- IFS detector window: the second surface of the singlet, with no optical power.

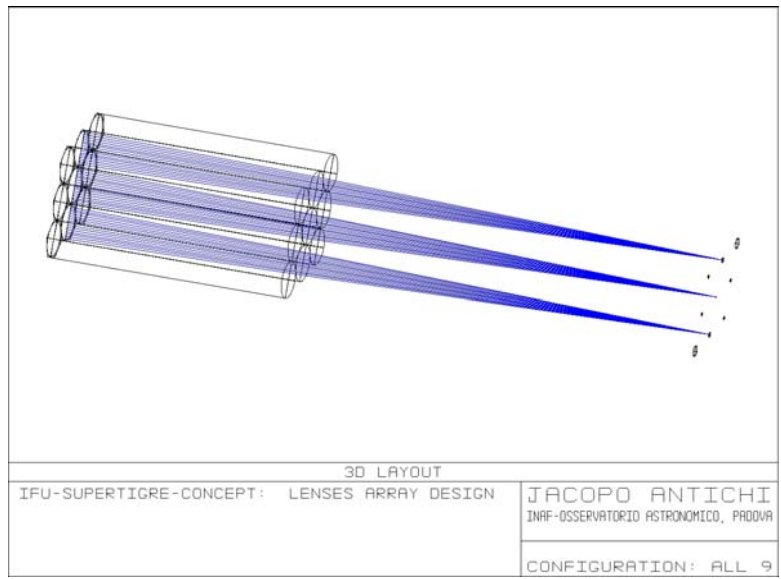

**Figure 4: Sketch of the IFU-system: the adopted glass is BK7; the pitch is 150 μm and the shape hexagonal.**

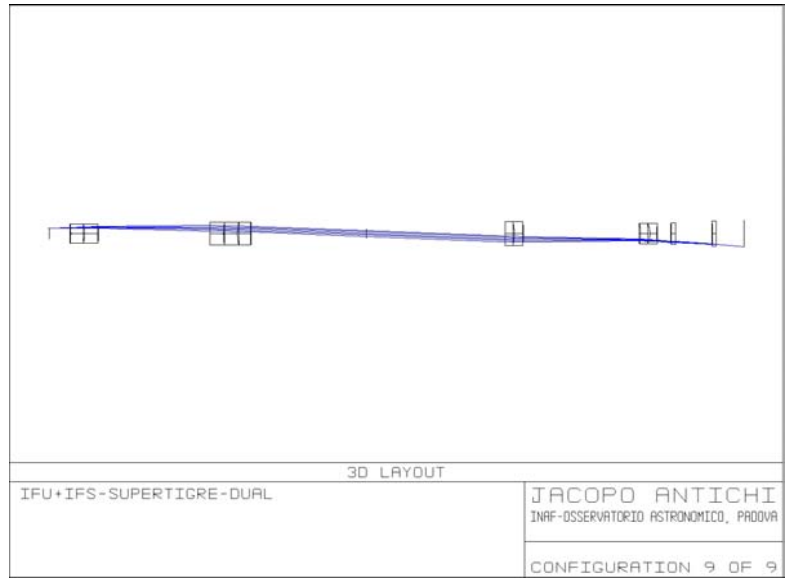

**Figure 5: Sketch of the IFS system: the adopted glasses are S-FTM16 and CAF2, the total length < 1.5 m.**

The system is optimized to work at the VLT temperature up to the detector window and with cryogenic conditions beyond this surface. The only cooled-optics is the field lens end the instrument background radiation is under control by imposing that the distance between detector window and detector pixels is > 150 mm.

## 4. CONTROLS

The IFS electronic control system is part of the SPHERE control system; it is based on a distributed network of VME-based local Control Units (LCUs) that manage the real time control, while a higher level of Unix based Workstations manage the user interface, the co-ordination, testing and maintenance.

The standard LCU is based on a stand-alone VME crate equipped with a Motorola CPU board, Ethernet board, and real time operating system VxWorks. Starting from this standard base, each LCU is equipped with an assortment of control and interface boards, depending on the specific functions. The equipment has been chosen in conformity with ESO VLT instrumentation standards, in order to ensure compatibility with other ESO applications and simplify maintenance and integration in VLT environment. The LCU uses the following boards:
- CPU board
- Motion controller boards
- DC motor amplifier boards
- Digital I/O board(s)
- Analog I/O Board
- Serial I/O board(s)

The main SPHERE-IFS subsystems to control are the motorized functions, instrument status and diagnostic, interlock system.

The motorized functions are based on the ESO standard Motion Control System, using DC servos with the correspondent couple of boards Maccon Controller/ESO Driver. This system will be used for general motion control applications. This standard set of boards is capable to drive 4 independent axes, and it is HW/SW configurable and can be adapted to different components.

Depending on the function, each implementation of the motorized function could be different in electro-mechanical properties and control requirements, as rotating/linear translation.

The Instrument Status and Diagnostic is a set of binary functions related to the status signals and alarm signals, and they come from relay contacts, hand push-buttons, proximity switches, temperature switches, etc. and the signal adaptation is realized by dedicated modules providing the galvanic isolation and the voltage adaptation for both input and output signals. They are then interfaced to the standardized Digital input/output VME Interface board. For the analogical measurements as temperature, current, voltage a stand alone data logger interfaced with the LCU via RS232 is used. In this way the monitoring the main parameters of the subsystem, is implemented enabling: a quick analysis of failures and malfunctions, the implementation of an historical database of the working parameters of the instrument.
The main parameters will be Moving Functions (MFs) and Electronic Cabinet (EC).
MFs:
- motor movements
- limit and Home switch status
- motor Current
- movements timeouts.

EC:
- ambient temperature
- surface temperature
- inner temperature

There are different signals in respect of the magnitude of the event. An out of range voltage will trigger a warning alarm, without any actions, but a fall of supply will trigger a fault alarm, with automatic actions (interlocks, stop motion…).
All the drive functions could be disabled using the terminals installed on the motor drive boards, essentially for the operational safety. Furthermore a interlock chain taking into account over temperature alarm, main stop function, and other conditions to be defined, is linked to the main interlock on VLT telescope.

## 5. DETECTORS

The IFS of SPHERE is designed to operate in the range 0.95-1.35 µm, i.e. where the most conspicuous broad molecular bands of the planets are expected to show off[9]. In these operational conditions our computations show that even keeping the optics at room temperature, the contribution of the thermal background to the noise can be neglected, if all radiation at wavelengths longer than 1.35 µm is blocked. This can be achieved either acting on the properties of the detector or inserting an edge filter. The cut-off wavelength of an HgCdTe detector for specific applications can be tuned by varying stoichiometric composition. However, this solution appears complex and unsatisfactory because of an excessively high read noise (~ 25 e$^-$) compared to standard detectors (~10 e$^-$). We have therefore investigated the other alternatives. With an edge filter transmitting about $10^{-4}$ at $\lambda > 1.35$ µm inserted in front of the detector and cooled down to LN2 temperature, the contribution of the thermal background to the detection noise can be neglected with respect to the other sources (photon noise of the parent star, RON, dark noise, etc.).

**Table 1: IFS detector requirements**

| Quantity | Specification | Goal |
|---|---|---|
| Size | 2048 x 2048 (≤20µm) | |
| DC | 0.1 e-/pix/s | 0.03 e-/pix/s |
| RON | ≤20 e- CDS | <15 e- CDS |
| Full Well | >100 ke- | <1000 ke- |
| Non Linearity(50%FW) | <1%(50%FW) - <3%(80%FW) | |
| Persistence | < 100 ppm | <100ppm |
| Charge Diffusion & CrossTalk | < 20% / 0.01% | |
| Cosmetics (Bad Pixels) | <2% | <1% |
| QE (90% of pixels) | >0.65 | ≥0.70 |
| PRNU (100 X 100) px$^2$ | < 5% RMS | < 1% RMS |

With such a solution a standard IR detector can be used for the IFS. Table 1 summarizes the main requirements of the IFS IR detector.

A crucial issue for planet finding is flat field accuracy. Taking into account that accurate flat fielding are foreseen and that images will be obtained using a dithering procedure that spread the signal of the planet over a large number of pixels, it is reasonable to assume that flat fielding accuracy of $10^{-4}$ can be achieved.

The detector should have pixel-to-pixel variations < 5% Root Mean Square (RMS) , and the goal is to have < 1% RMS in the whole wavelength range (values estimated excluding defect pixels). This value of PRNU is not required all over the detector but in areas of 100x100 pixel. This means the requirement on PRNU on larger scales can be relaxed up to <10%.

## 6. CRYOSTAT

Figure 6 shows two the current configuration of the IFS cryostat in the case LN2 is adopted to cool down the various part of it. As one can see the LN2 tank is placed vertically respect to the optically bench.

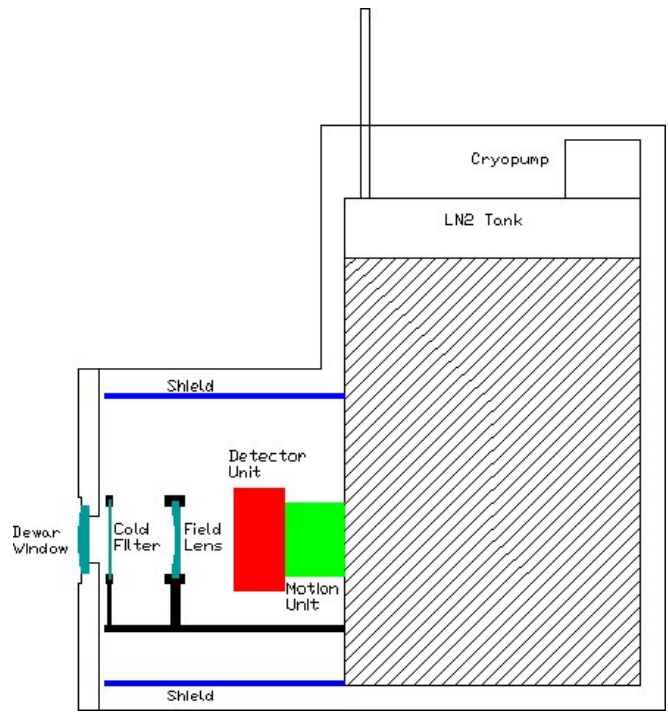

**Figure 6: A sketch of the IFS cryostat in which the LN2 tank is perpendicular to the direction of the optical path**

The cryostat will be composed by the following parts:
- The Detector Unit: comprising the detector itself, the detector mount and the detector front end electronics.
- The Motion Unit: comprising the cryo-mechanisms for dithering and the detector unit holder
- Optical Unit: comprising the cold filter, the field lens and the optical bench.
- Cryostat Flange comprising the vacuum sealed flange and the cryostat window.
- Thermal Shields.
- LN2 Tank.
- Temperature control unit.

## 7. SOFTWARE

The IFS instrument software (INS) is part of the SPHERE software, whose concept has been developed according to the general standard for ESO VLT instrument software[10]. The instrument control system is therefore made up of a distributed network of local microprocessors (Local Control Unit, LCU), which control the instrument hardware functions, and Instrument Workstations (IWS), devoted to subsystems coordination and monitoring.

The IFS INS consists of:
- the Instrument Control Software (ICS), which controls all instrument functions (except for the detector) and monitors all sensors;
- the Detector Control Software (DCS), which controls all functions belonging to the detector sub-system and transfers detector data to the Instrument WS, for the IFS the standard IRACE controller and software will be used[11];
- the Observation Software (OS), which is in charge of the coordination of all IFS subsystems.

Observations are specified by means of Observation Blocks (OBs) that are delivered by the High-level Observation Software (HOS) to the Broker of Observation Blocks (BOB) for execution. Commands are then sent to the SPHERE Super OS (SOS) that coordinates all software the control the instrument subsystems (Adaptive Optics, Zimpol, IFS) and the telescope (TCS).

Because of the way observations are performed in the IFS (i.e. SSDI), it is required to have an on-line, possibly partial, data reduction pipeline for quick-look of data and quality control. This pipeline will optimally extract and calibrate the spectra from the data, produce monochromatic images, estimate the image quality, then rotate and sum to integral images in case the quality test is passed, and visualize the images and data quality parameters within a duty cycle (i.e. a few sec). Since such kind of online processing is currently not supported in the VLT SW, it will be the main subject of study during the preliminary design of the IFS INS in order to devise a solution that fulfils requirements while remaining "VLT compliant" as much as possible.